# Automatic Diagnosis of Pneumothorax from Chest Radiographs: A Systematic Literature Review

Tahira Iqbal, Arslan Shaukat, Usman Akram, Zartasha Mustansar
National University of Sciences and Technology (NUST), H-12, Islamabad, Pakistan
tahira.iqbal36@ce.ceme.edu.pk, arslanshaukat@ceme.nust.edu.pk, usman.akram@ceme.nust.edu.pk; zmustansar@rcms.nust.edu.pk

## Abstract:

Among various medical imaging tools, chest radiographs are the most important and widely used diagnostic tool for detection of thoracic pathologies. Research is being carried out in order to propose robust automatic diagnostic tool for detection of pathologies from chest radiographs. Artificial Intelligence techniques especially deep learning methodologies have found to be giving promising results in automating the field of medicine. Lot of research has been done for automatic and fast detection of pneumothorax from chest radiographs while proposing several frameworks based on artificial intelligence and machine learning techniques. This study summarizes the existing literature for the automatic detection of pneumothorax from chest x-rays along with describing the available chest radiographs datasets. The comparative analysis of the literature is also provided in terms of goodness. Limitations of the existing literature along with the research gaps is also given for further investigation. The paper provides a brief overview of the present work for pneumothorax detection for helping the researchers in selection of optimal approach for future research.

## Introduction

Chest radiography is an economic and commonly used medical imaging technique for diagnostic purpose. Moreover because of the cheap cost and availability of X-ray machines in almost everywhere, doctors prefer X-rays instead of other medical imaging techniques like Magnetic Resonance Image (MRI) or computed tomography (CT) scan. Chest X-rays (CXR) are widely used for automatic diagnosis of lung diseases including pulmonary nodules, tuberculosis and other lung diseases [1]. One of the common chest pathologies is pneumothorax which can be life threatening if not treated in time. Almost 30-39% patients suffering from chest trauma have chances to develop pneumothorax as well [2]. Pneumothorax also known as Collapsed Lung refers to a condition in which the air present inside the lung is leaked into the pleural cavity (cavity or empty space between lungs and the chest walls). This air exerts pressure on the lungs which results in collapse of lung. The most commonly observed symptoms include sudden, sharp pain in chest and shortness of breath. It occur because of several reasons such as injury in chest area or some other lung diseases, however sometimes there is no apparent reason for the lung collapse [3]. The size of affected area by pneumothorax vary from small to large and it is the size of the collapsed lung which is decisive factor for the treatment. However because of the complex overlapping structure of thoracic cavity and the fact that the position and size of pneumothorax vary from patient to patient, it is a challenging task to diagnose the disease [4]. Moreover because of less number of radiologists throughout the world, a single radiologist has to analyze thousands of X-rays per year [5]. Hence in order to lessen the burden on the radiologist and to provide him a second opinion while analyzing a radiograph, a Computer Aided Design (CAD) can be a game changer and it has become a topic of interest in the field of chest radiography.

The very first CAD system was proposed in 1960 and the results proved that the diagnostic performance was increased by a CAD system [6]. In medical field CAD systems have been used for several purposes including arrhythmia detection [7], classification of skin cancer [8] and diabetic retinopathy identification [9]. Any CAD system comprise of four main steps. (1) Preprocessing of the data (2) extraction of region of interest. (3) Feature extraction step, which can be done using different ways like Local binary pattern, textural feature extraction. (4) Classification of the input into respective class based on the features. The last step need a classifier to be trained on some data, (which are the features extracted from the input data). Different machine learning classifiers include Linear Regression, Support Vector Machine, Naive Bayes Classifier, Radom Forest or Neural Networks [10], [11].

Before the advent of deep learning techniques, different thoracic pathologies have been successfully detected using traditional machine learning approaches [12]. However it has few drawbacks, firstly the features are handcrafted so the choice of optimal feature selection method is an important decision to be made. Secondly a machine learning classifier need to be selected depending on the resources available. Hence this process is time consuming. On contrary the processes of feature extraction and classification have been automated by deep learning techniques. The design of Artificial Neural Networks was aspired from the working mechanism of human brain [13]. Convolutional Neural Networks have achieved astonishing results in automating different process in different fields of life including agricultural domain [14] surveillance [15] and medical field [16].The only drawback of deep learning methods is the dire need of enormous data for training purpose, which is not easy to acquire and it is time consuming to label and annotate unlimited number of samples.

With the passage of time, datasets are being made available and people are putting their efforts in providing large annotated datasets. Kaggle is one of the most commonly used and free platform where almost 70,000 different types of datasets are available which can be utilized for research purpose [17]. Moreover Google Dataset Search provides a facility to search all the available datasets related to any topic [18]. The availability of Chest x-rays datasets have allowed researchers to contribute for automating the chest diseases detection. However one of the common issue with medical images datasets is the class imbalance problem, i.e. the number of samples for each class is not the same [19]. This issues is resolved in different researches by using various techniques including data level and algorithm level techniques [20].

To be precise, in the field of automatic diagnosis of thoracic pathologies, different CAD systems have been proposed which found their roots from deep learning. Examples include lung nodule detection [21], detection of pneumonia and tuberculosis [22], [23] and diagnosis of other lung diseases including pneumothorax [24]. Recently, U-Net, a segmentation model evolved from convolutional layers was proposed which is very commonly used for the purpose of medical imaging segmentation [25]. This paper presents the overview of the existing work for pneumothorax detection using Chest radiographs. The usability, advantages and drawbacks of the techniques utilized is also analyzed and discussed along with a brief introduction to the available CXRs datasets.

The remaining papers is organized as follows. Section 2 briefly talks about the research methodology used for the systematic literature review. The different openly available Chest X-rays datasets are mentioned in detail in Section 3. In Section 4 the techniques proposed by researchers for pneumothorax detection are discussed. Section 5 gives the overview of the evaluation metrics used in most of the literature and Section 6 provide comparative analysis of the existing work. In Section 7 conclusion is drawn based on the entire review.

## 2. Research Method

An unbiased research methodology is adopted for a systematic literature review (SLR) in order to ensure the evaluation of all the existing research relevant to the said field.

### 2.1 Data Sources

Relevant studies have been obtained mainly from five electronic databases which are enlisted in Table 1. The total number of search results in Table 1 shows the number of records after applying the Selection criterion as mentioned in section 2.3.

**Table 1**: List of Available Databases

| Databases | Link | Search Results |
|---|---|---|
| Springer | https://link.springer.com/ | 248 |
| Science Direct | https://www.sciencedirect.com/ | 184 |
| arXiv | https://arxiv.org/ | 117 |
| IEEE | https://ieeexplore.ieee.org/ | 103 |
| Nature | https://www.nature.com/ | 21 |
| Hindawi | https://www.hindawi.com/ | 5 |

### 2.2 Search Terms

Certain search terms have been utilized in order to search the relevant literature and the valuable work done in the field of automatic diagnosis of pneumothorax.

i. Pneumothorax OR pneumothorax detection using artificial intelligence
ii. Diagnosis of chest pathologies OR pneumothorax from Chest radiographs
iii. Chest X-rays datasets OR datasets for pneumothorax identification
iv. Mentioned below is the search string for the said purpose:
pneumothorax OR using OR chest OR x-rays OR artificial OR intelligence

### 2.3 Study selection procedure

Some of the criterion are adapted in order to select the relevant studies out of all the literature retrieved from the data sources. These inclusion and exclusion criteria are briefly explained below:

**Inclusion criteria:**

- Studies relevant to pneumothorax detection by using chest radiographs.
- Studies that were not just specified to pneumothorax, instead covered multiple chest diseases detection including pneumothorax.
- Studies utilizing public and private datasets.

- Studies published in peer reviewed publications along pre-prints (arXiv).
- Studies that were published from 2010 to 2020.
- Studies using English language as mode of sharing the research.

**Exclusion criteria:**

- Studies which did not add value in the field of automatic diagnosis of pneumothorax.
- Studies relevant to pneumothorax using chest radiography reports in textual format.
- Studies relevant to pneumothorax using medical imaging techniques other than X-rays.
- Studies published in language other than English.
- Web articles, Wikipedia, posters and short papers.

## 3. Datasets

The openly available dataset which can be utilized for automatically detecting the presence of pneumothorax are summarized in this section.

### 3.1. NIH Chest Xray14 dataset

A very large chest x-rays dataset was presented by Wang in year 2017 [26] containing 112,120 CXRs from 30,805 patients and named as Chest X-ray14 dataset [27]. Picture Archiving and Communication Systems (PACS) of the hospitals affiliated with National Institutes of Health Clinical Center (NIH) were used for acquiring the chest radiographs. This dataset initially covered 8 thoracic pathologies but later it was extended to 14 chest diseases including atelectasis, edema, consolidation, emphysema, cardiomegaly, effusion, mass, fibrosis, infiltration, hernia, nodule, pneumothorax, pleural thickening and pneumonia. Out of 112,120 CXRs, 60,412 samples are those with no-pathology while remaining 51,708 CXRs have one or more pathologies. In the whole dataset 5298 CXRs are labelled to have pneumothorax. Moreover the dataset also provides bounding boxes information which can be used for disease localization however this information is for only 980 CXR images .The resolution of each CXR in this dataset is 1024 x 1024. In addition to CXR images, the metadata, e.g. patient gender, age, and view position of the CXR are also available. Training and testing lists are also provided in this dataset in which the samples were split in such a manner that CXRs from same patient belong to either training or testing list. Some of the images NIH CXR14 are shown in Figure 1.

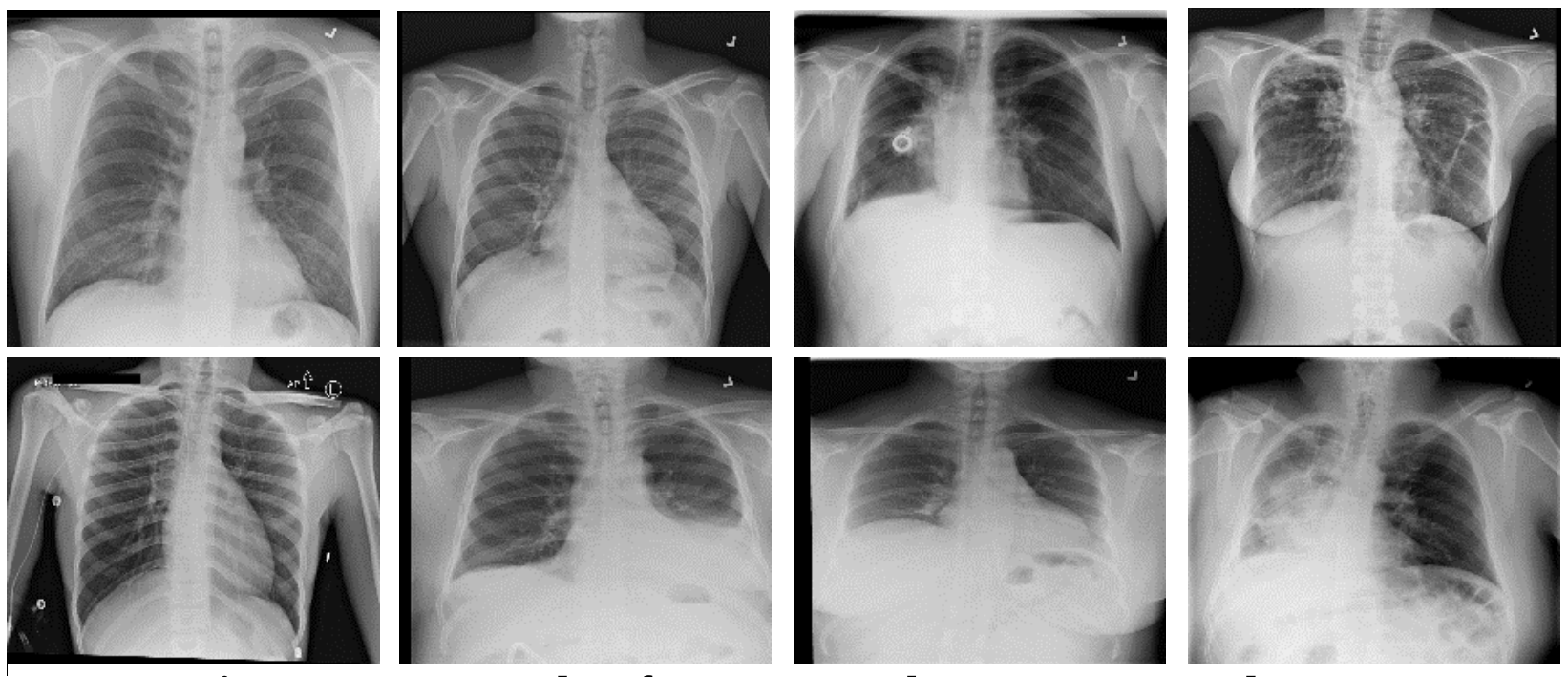

**Figure 1:** Samples from NIH Chest X-ray-14 dataset

### 3.2. Random Sample of NIH Chest X-ray dataset (RS-NIH)

A small version of NIH Chest X-ray14 (NIH CXR14) dataset is available on Kaggle [28] and contains 5% of whole dataset which were randomly selected. The samples were selected in such a way that the percentage of occurrence of each pathology is the same as the NIH Chest Xray14 dataset. Total of 5606 CXRs are present covering 14 thoracic pathologies and 15$^{th}$ class being the samples with no-pathology. Just like in NIH CXR14 dataset, the resolution of each image in this dataset is 1024 x 1024. Among 5606 CXRs, 3044 images are of healthy persons while remaining CXRs correspond to one or more thoracic diseases. Meta data like patient age, patient gender, patient's follow up, view position etc. are also supplied for each CXR image in the database. Some of the images belonging to RS-NIH are displayed in Figure 2.

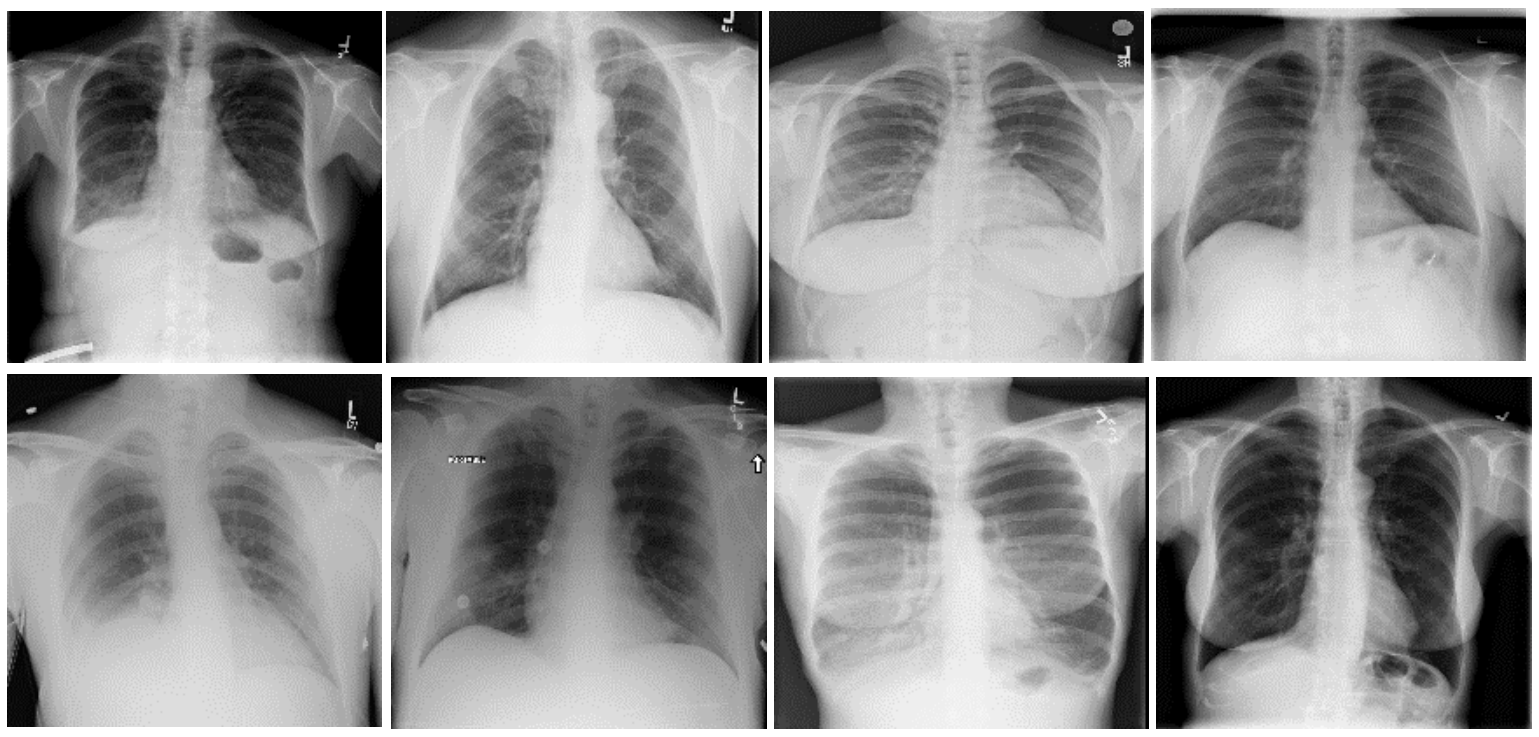

**Figure 2:** Examples taken from RS-NIH dataset

### 3.3. CheXpert

A large openly available dataset was made available in 2019 by the Stanford group [29] [30]. It contains 224,316 Chest x-rays of 65,420 persons obtained from Stanford Hospital, collected between 2002 to 2017.It covers 14 common chest diseases including atelectasis, enlarged cardamom, lung lesion, cardiomegaly, edema, fracture, lung opacity, pneumonia, consolidation, pleural effusion, pneumothorax, pleural other, no finding and support devices. The typical size of each CXR is 320 x 320 in this dataset. A unique feature of this dataset is that it provides uncertainty label as well for every pathology, i.e. those CXRs in which the labeler couldn't properly interpret regarding presence or absence of certain pathology is stated to be uncertain for a particular pathology. These uncertainty labels can be dealt with different techniques [29], [31]. The dataset contain 16627 CXRs with no pathology while remaining CXRs have either one or more chest pathologies. 17313 samples belongs to pneumothorax class while there is uncertainty regarding the declaration of 2663 CXRs to have pneumothorax. Some of the images from this dataset are shown in Figure 3.

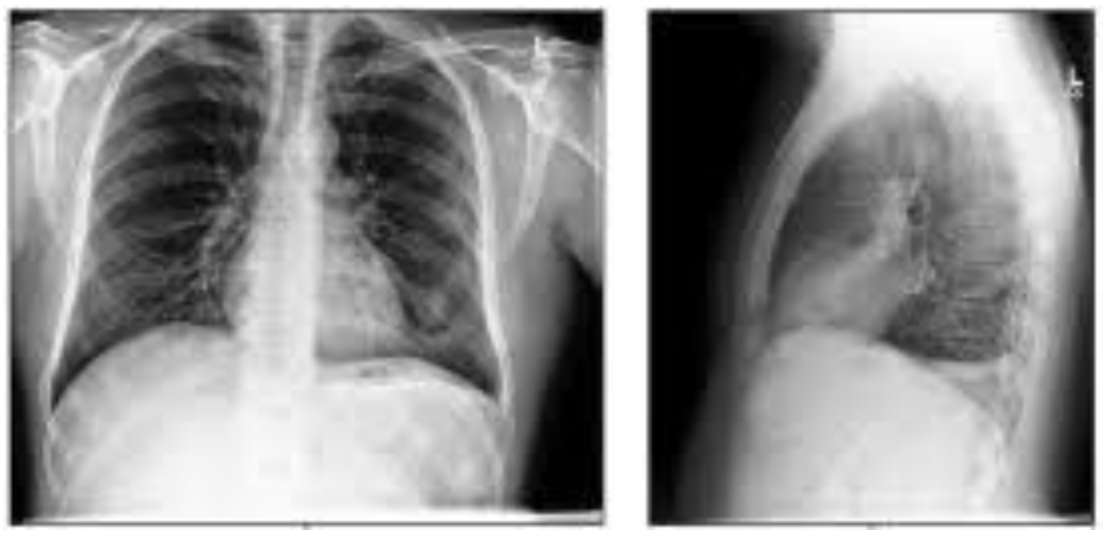

**Figure 3:** Images from CheXpert dataset

### 3.4. MIMIC-CXR-JPG

To date, it is the largest openly available Chest X-rays dataset, which was made public in 2019 [32], [33]. It contains 377,110 CXRs which are related to 22,7827 radiologic studies and covers 14 chest diseases including No Finding, atelectasis, cardiomegaly, edema, consolidation, enlarged Cardio mediastinum, lung lesion, fracture, lung opacity, pleural Effusion, pneumothorax, pneumonia, pleural Other and support Devices. The CXRs were acquired from Beth Israel Deaconess Medical Centre (Massachusetts, USA) in the duration of 5 years (2011 to 2016). 75163 images are of the people with no pathology while remaining CXRs belong to the patients who have one or multiple thoracic pathologies. There are 9317 CXR images labelled as pneumothorax in the whole dataset while for 868 images, there is uncertainty regarding presence of pneumothorax. Some of the sample images from MIMIC-CXR-JPG dataset are shown in Figure 4.

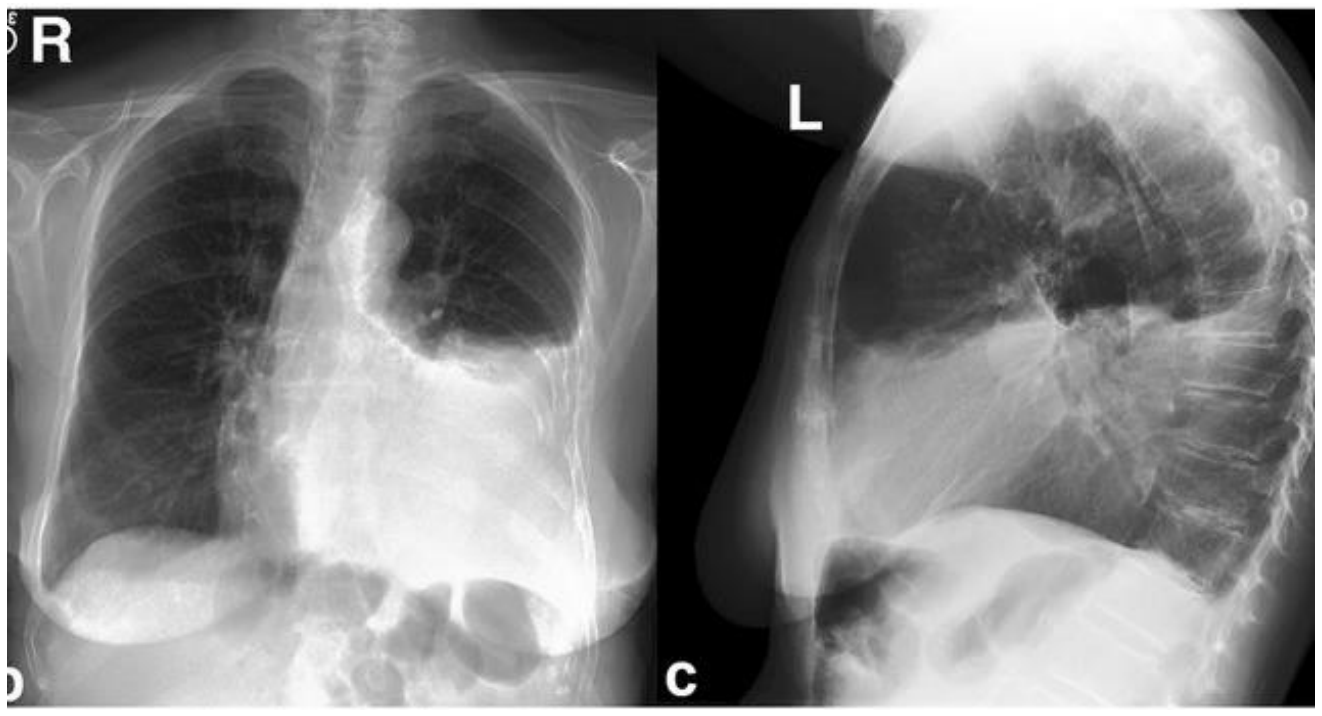

**Figure 4**: Samples from MIMIC-CXR-JPG dataset

### 3.5. SIIM-ACR Pneumothorax Segmentation

A very recent dataset provided by Society for Informatics in Medicine (SIIM) in 2019 is a very major contribution in the field of automated detection and localization of pneumothorax [34]. The images were acquired from NIH Chest Xray14 dataset which were later annotated on pixel level using artificial intelligence techniques and reviewed by expert radiologists. The dataset is available on Cloud Healthcare[35] [36] in which the CXR images are provided in DICOM format while corresponding masks for each image are in Run-Length Encoding (RLE format) format, however many individuals have shared the same dataset in PNG format on Kaggle [37], [38]. Since this dataset was released as challenge/ competition and invited people to participate in developing robust segmentation model, thus it has two stages, the stage1 has 10,675 samples for training purpose and 1372 samples belong to test set, while in stage2 the number of samples in both training and testing set was increased to 12,047 and 3205 samples respectively. Most of the researchers utilizing this dataset have evaluated their proposed models on these held-out test sets (i.e. stage1 test set and stage2 test set).  Few images from this dataset are shown in Figure 5.

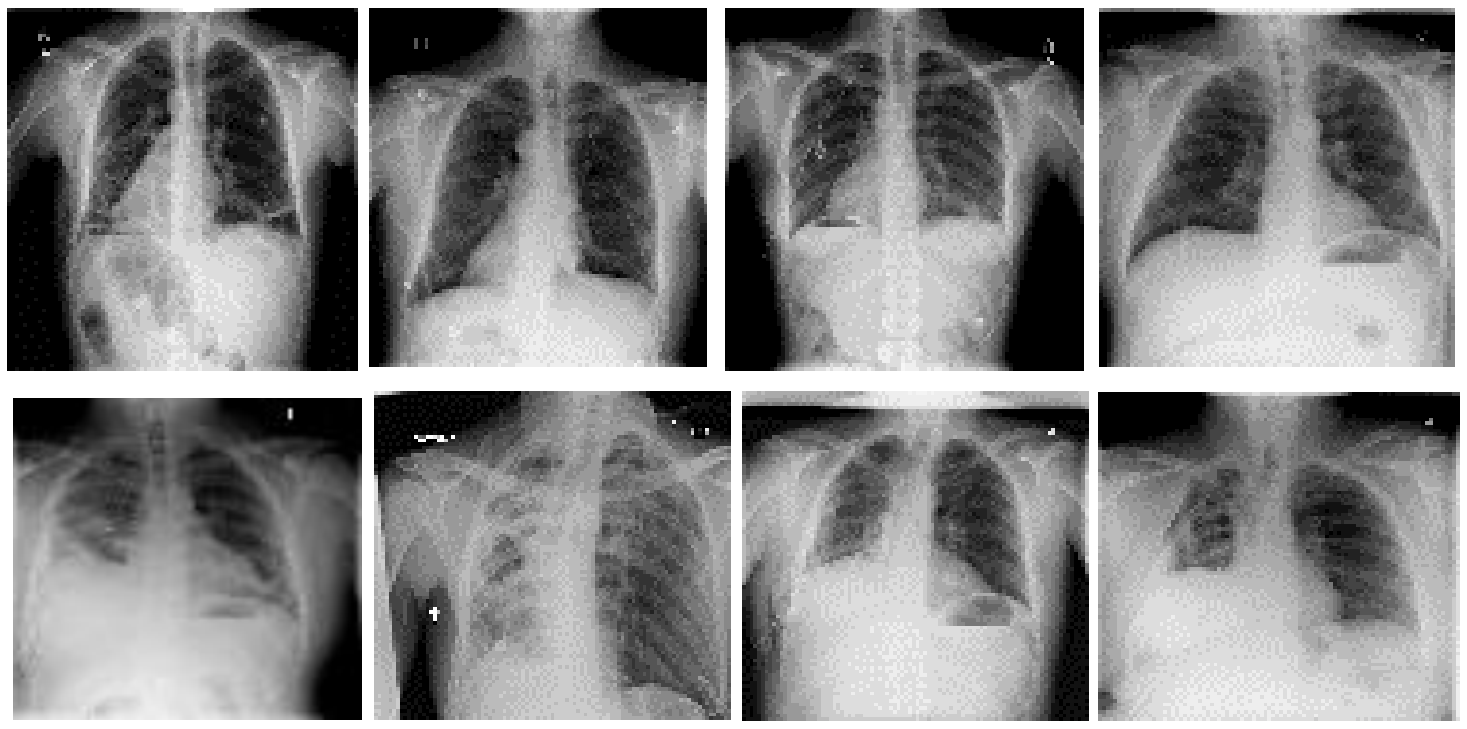

**Figure 5:** Samples from SIIM ACR dataset

The total number of CXRs labelled as pneumothorax in each of the above mentioned dataset are stated in tabular form in the Table 2. The Table 2 also highlights if the dataset provides data-splits for training and testing purpose or not.

**Table 2.** Summary of databases with respect to number of pneumothorax samples

| **Year** | **Dataset** | **Total No of samples** | **No. of Pnu samples** | **No. of class in dataset** | **Official Data split** |
|---|---|---|---|---|---|
| 2017 | NIH Chest X-ray14 dataset [27] | 112,120 | 5298 | 14 | ✓ |
| 2017 | Random Sample of NIH Chest X-ray dataset [28] | 5606 | 271 | 14 | ✗ |
| 2019 | CheXpert [29] | 224,316 | 17,313 | 14 | ✗ |
| 2019 | MIMIC-CXR-JPG [33] | 377,110 | 9317 | 14 | ✗ |
| 2019 | SIIM ACR Pneumothorax Segmentation [36] | 10,675 (stage1) 12,047 (stage2) | 2379 (stage1) 2669 (stage2) | 2 | ✓ |

Pnu: Pneumothorax

## 4. Pneumothorax detection methodologies

This section is divided into two main categories. Firstly the research done exclusively for automatic detection of pneumothorax is discussed followed by the existing valuable literature covering multiple pathologies detection including pneumothorax. The first category is further split into three sub-categories: 1. Classification (i.e. publications in which main focus was to label a CXR as either normal or pneumothorax). 2. Localization (papers in which the main goal was to achieve the area of pneumothorax). 3. Classification and Localization (research which covers both classification of CXR as either normal or pneumothorax and identification of region of pathology).

On the other hand, second category is split into two categories, which are (1) Classification and (2) Classification and Localization.

### 4.1. Pathology specific research.

#### 4.1.1. Classification

Jun [39] proposed an ensemble based framework in order to distinguish between normal and pneumothorax CXRs. The ensemble model comprised of three identical CNN architectures trained on three different input sizes. Pretrained ResNet50 [40] architecture was used which was trained on same training set with different resolutions, i.e. 512 x 512, 384 x384 and 256x256. ImageNet weights were used for model initialization and RMSprop optimizer was selected for model training. The ensemble was created by averaging the probabilities generated by each of the trained model. NIH Chest X-ray14 dataset (NIH) was used in this experiment which contains 112,120 images however for this study 59,156 samples of Normal CXRs and 5225 samples of pneumothorax were used. These 64,381 CXRs were randomly divided into 80-20 for training and testing purpose respectively. To avoid the problem of class imbalance, instead of directly using the probability from the softmax layer, a cut-off value was selected so as to maximize the sum of specificity and sensitivity in ROC curve. When evaluated on the test set, the ensemble model achieved AUC of 91.1% for pneumothorax detection.

Sze [41] proposed a pneumothorax detection model in which the power of transfer learning was explored. A 122 layered deep neural network was proposed which was named as tCheXNet based on the fact that it utilized the trained model CheXNet [22] in order to initialize the model weights. CheXpert dataset was selected for this experiment. The model was trained using 94,948 CXRs in which only 13911 CXRs belonged to pneumothorax class, this class- imbalance problem was solved by using weighted binary cross entropy loss function. The proposed framework was tested on a test set containing 7 Pneumothorax images and 195 Normal CXR and achieved AUC of 70.8%.

In [42], an exclusive framework was proposed for automated diagnosis of pneumothorax from the chest x-ray images by combining the texture analysis method with supervised learning techniques. Two main steps were involved in the proposed framework. (1) Detection of abnormal texture of the lung caused by pneumothorax, using texture analysis process. Two types of textural feature extraction approaches including Local Binary Pattern (LBP) and Maximum Response Filters (MR) were experimented in this research. This way labeled patches of images were generated after which the local analysis values were integrated with global image representation. (2) The global images were then indulge into training purpose for identifying the presence of pathology. The global images used for training were designed on the basis of shape of the lung. Both local and global data was incorporated for supervised learning process. Gentle AdaBoost classifier and KNN classifiers were used for the training purpose. Local dataset was used in this research which was obtained from Sheba Medical Center and it comprised of total 108 cases out of which 48 CXRs were diagnosed with pneumothorax. When evaluated on the test set the proposed model achieved 81% sensitivity and 87% specificity value.

Taylor [43] collected Chest X-rays images from Center for Digital Health Innovation at UCSF in order to design a model for automatic detection of pneumothorax. The dataset contained 13,292 CXRs belonging to two classes, pneumothorax and No-pathology. The images were in DICOM format which were converted to JPEG and resized to 512x512. The dataset was split into 70% training, 15% validation and 15% testing set. The training set had the issue of class imbalance as it contained 2214 positive and 7095 negative samples so this issue was resolved using random under-sampling technique. Image Augmentation was also used in order to avoid overfitting. Training with several CNN architectures including VGG-16, VGG19 [44], ResNet, Inception [45] and Xception [46] was carried out in this study. Performance was assessed on the internal and external testing set which was obtained from NIH Chest Xray14 dataset. The results showed that there was a performance decline in case of external test set. For the internal test set the proposed framework attained AUC of 94% with sensitivity of 55% and specificity of 90% while on external test set AUC of 75% with 49% sensitivity and 85% specificity was achieved.

In [47] a framework evolved from deep learning techniques was proposed for diagnosis of pneumothorax. The classification model was named as ChestNet in which data cleaning was done using two Network in Network (NIN) and a new data augmentation method based on random histogram equalization was introduced for data imbalance problem. The training and evaluation of the proposed model was done on two datasets including ZJU-2 and NIH Chest Xray14 dataset. The ZJU-2 dataset was obtained from hospital affiliated with Zhejiang University School of Medicine China. The CXRs were randomly divided into 80% training and 20% test set. For training purpose the samples of ZJU-2 and NIH were combined together. The training set contained 26891 non-pneumothorax samples, 1507 and 5298 pneumothorax samples from ZJU-2 and NIH respectively. The input resolution was kept 480 x 480 and the proposed model achieved AUC of 98.44% on ZJU-2 and 99.06% AUC for NIH Chest xray14 test set. Different visualization techniques were also studied including thermogram and texture visualization however the performance for the visualization techniques was not reported. Thermodynamic images were generated by using the GradCAM method in order to visualize the area of pathology while the texture features learnt via Guided Backprop method showed the rib texture of thoracic region.

A study was carried out by Park et al [48] in order to estimate the performance of CNN for the automatic diagnosis of pneumothorax from the CXRs after Percutaneous transthoracic needle biopsy (PTNB). PTNB is a famous and widely used technique for detection of pulmonary lesions. The data was obtained from Picture Archiving and Communication System (PACs) of two local hospitals, and it consisted of 1596 CXRs with pneumothorax and 11137 normal CXRs. During training, 1343 pneumothorax CXRs were used, out of which 90% samples were randomly assigned to training set and remaining 10% were treated as validation set, while a held-out test set containing 250 normal cases and 253 images with pneumothorax were used for internal testing purpose. Pretrained YOLO Darknet-19 [49] was fined tuned on the CXRs dataset with SGD optimizer and initial learning rate was set to 0.001. The input resolution was kept as 1024 x 1024. Contrast Limited Adaptive Histogram Equalization was used for preprocessing. The trained model achieved sensitivity of 89.7% and specificity of 96.4% with AUC of 98.4% on the internal test set. The model was also tested on an external validation set containing 309 pneumothorax and 1020 normal CXRs and achieved sensitivity, specificity and AUC of 63.4%, 93.5% and 90.5% respectively.

#### 4.1.2. Localization

Mostayed [50] suggested a modification for the traditional U-Net architecture [25] which is known for its stand-out performance for extracting the region of interest from medical images. The modification was proposed with the aim to lessen the network’s trainable parameters by replacing the concatenation operation in the skip connection of traditional U-Net architecture with content-adaptive convolution. The proposed method was experimented with two different datasets, one of which was SIIM Pneumothorax dataset. 12,047 CXRS were utilized in this research which were split into training and validation set. The input resolution of images was kept equal to 128x128. RMSProp was selected as optimizer and initial learning rate was set to 0.001. A customized loss function was defined by combining binary cross entropy and soft dice loss. The segmentation score for both the traditional and modified U-Net were reported for the test set along with the number of trainable parameters for both type of networks. The U-Net model achieved dice coefficient score of 75.36% with 7.76M parameters while the modified U-Net achieved 76.04% dice coefficient score with 7.09M parameters. Hence the modified U-Net not only increased the dice coefficient score but also reduced the resources utilization.

In [51], a segmentation model was proposed for the precise prediction of the portion of lung affected by pneumothorax and generated a binary mask to assist the radiologist in identifying the location and size of pneumothorax. U-Net architecture was used in this study in which the traditional encoder part was replaced with ResNet model. Apart from resizing the input images from 1024 x 1024 to 256 x 256, contrast correction was also done in order to have a uniform color range in the image. SIIM Pneumothorax dataset was used for experiment purpose containing 12,047 CXRs. 80% of the dataset was used for training purpose while remaining 20% was used for validation and testing purpose. During training of the U-Net architecture, initial learning rate of 0.0001 with Adam optimizer was used. For better training, binary cross-entropy loss function and early stopping callback were employed. The model performance was evaluated in terms of Dice Coefficient score (DSC) and Intersection over Union (IoU) which was reported to be 84.3% and 82.6% respectively.

In [52], a weakly supervised segmentation model was proposed with the aim to reduce the dependency on pixel level annotation and utilizing only image-level labels. The proposed method comprised of three consecutive steps. (1) Class Activation Maps generation was done by training ResNet50 architecture using the image-level annotations of the training data and then GradCAM++ method was utilized in order to generate the activation maps from the trained model. (2) These generated activation maps were trained on IRNet, which is known for its ability to improve the inter-object bounders. This step was performed with the aim of improving the quality of these maps prior to segmentation task. (3) Finally the output from the second step was trained on U-Net model with ResNet50 backbone. Experiments were performed on two different datasets, one of which was SIIM Pneumothorax dataset using the training set of stage 1 containing 2379 samples with pneumothorax and 8296 normal cases. However the stage1 test set was further split into validation and test set, thus validation and test had 145 positive and 541 negative cases each. Weighted binary cross entropy (BCE) was used for avoiding the class imbalance issue, while overfitting problem was reduced by data augmentation method. SGD was used as optimizer with initial learning rate of 6e-5 and momentum was set as 0.9. The image resolution was kept 512 x 512 with a batch size of 48. The model evaluation was done on test set which achieved dice coefficient score of 76.69%.

Abedalla [53] proposed a two stage segmentation framework (2ST-Unet) in order to extract the region of pathology. The main architectural block was U-Net with ResNet34 encoder, while the weights of the model trained on ImageNet dataset were used to initialize the encoder part. The decoder block comprised of five blocks containing several convolutional and batch normalization layers and RELU activation function layer. The first phase was trained with image resolution of 256 x 256 and then the weights from this phase were used to initialize the training of second phase for which image resolution was kept 512x 512. For model training, Adam optimizer was used and initial learning rate was set to be 0.001 which was decayed by using cosine annealing scheduling method with every epoch. Stochastic Weight Averaging SWA was also applied to last few epochs for fast convergence Different data augmentation techniques were also applied in order to avoid overfitting. SIIM Pneumothorax dataset was used for this research with the officially provided data split for training and testing set. 2669 pneumothorax and 9378 normal CXRs are present in the training set. The evaluation was performed on the test set of stage1 and stage 2 separately containing 1372 and 3205 samples respectively. Test Time augmentation was also applied and the trained model achieved Dice coefficient score equal to 85.02% for stage1 and 83.56% for stage 2 test set.

Tolkachev [54] proposed a segmentation model for identifying the area of lung affected by pneumothorax using SIIM Pneumothorax dataset containing 12,047 CXRs. A U-Net model was used in which various CNNs including SE-ResNext50, ResNet34, DenseNet121 and SE-ResNext101 were used to replace the traditional encoder. In addition to resizing the images from 1024 x 1024 to 768 x 768, some other pre-processing was also applied before training the segmentation model. In order to enhance the model performance, different data augmentation techniques were also exploited. The encoder part of the U-Net architecture was initialized with ImageNet weights while binary cross entropy was used as a loss function. Stratified five-fold cross validation method was adopted during training process. The performance of the proposed model was evaluated on a test set containing 1372 CXRs for which it achieved dice coefficient score of 85.74%.

A major contribution in the field of localization of the area of pneumothorax was made in [55]. Here an ensemble of three LinkNet networks [56] was used with three different backbones including SENet154, seresnext50, seresnext101. The original encoder part of the LinkNet was replaced by these backbone architectures. The decoder block consisted of convolutional layer followed by an up-sampling and a convolution layer. After every layer there was a batch normalization and RELU activation layer. Lastly sigmoid activation was used for acquiring the probability values. The training for each of the LinkNet model comprised of three phases. (1) Training for 40 epochs with Adam optimizer and Cosine Annealing scheduling. (2) Training for 15 epochs with CyclicLR scheduling. (3) Training using the initial setup for 15 epochs. Data augmentation during training and Test Time Augmentation was incorporated in this research. Since the dataset was highly class-imbalance, so in order to ensure the presence of positive samples in every batch, "non-empty sapling" was performed. The ensemble of the three models was created not only by traditional method of averaging the probability values but also union of binary values of the predicted masks was taken. The experiment was done using SIIM Pneumothorax dataset. 10675 samples were present in training set while the evaluation was done using stage1 and stage2 test set separately. The original resolution of images (1024 x1024) was used without any preprocessing with a small batch size 2. The proposed ensemble model achieved Dice coefficient score of 88.21% and 86.14% for stage1 and stage2 test set respectively.

A novel segmentation framework was proposed for identifying the location of pneumothorax in [57]. Three modules were combined in the proposed framework including Fully Convolutional DenseNet, scSE which is a spatial and channel squeeze and excitation module and finally a multi-scale module. Single channel images which were resized to 256x256 resolution were used in this research. The dataset was obtained from the institutions' PACs. The dataset contained 11051 frontal CXR images out of which 5566 belonged to Pneumothorax and 5485 were the x-rays of healthy persons. The data was split into training, validation and test set. For solving the pixel-level-class-imbalance, weights were added in the cross entropy function and named as weighted cross entropy loss (W-CEL). Adam optimizer was adopted with initial learning rate set to 1 e - 4. Moreover data augmentation by means of horizontal flip was carried out for the whole training set. The trained model achieved Dice coefficient score of 92.01% with Mean pixel accuracy (MPA) of 93.01% when evaluated on the test set containing 2213 CXRs.

In [58] a deep learning based framework was proposed for automatically segment out the region of pathology. The main idea behind this framework was to lessen the dependency on the pixel-level annotations for segmentation purpose. The two main stages were (1) training an image-level classifier which not only predicted the class of the CXR images but also generated the attention masks, these attention masks gave a rough estimation of the area of pathology. These masks had some errors which were corrected using SLSR (spatial label smoothing regularization) technique. (2) Segmentation model was trained using some of the image which were well annotated on pixel level along with the weakly annotated masks generated by the image level classifier. The Image level classification was performed using ResNet101 model and the weakly annotated masks were obtained by utilizing Guided Attention Inference Network (GAIN). For segmentation purpose, three different type of architectures including U-Net, LinkNet and Tiramisu (FCDenseNet67) [59] were experimented while keeping input resolution of 256 x 256. The experimental data was obtained from Mianyang Central Hospital, Mianyang, Sichuan China and it contained total of 5400 CXRS with 3400 pneumothorax case and 2000 normal CXRs. Out of 3400 CXRs, only 800 pneumothorax cases were well annotated in terms of pixel level annotations (i.e. ground truth mask). The dataset was evenly and randomly divided into training and testing set, thus training set comprised of 4000 CXRS while 1400 CXRs were present in test set. When evaluated on test set, the model performance was reported 66.69% in terms of IoU score.

#### 4.1.3. Classification and Localization

Machine learning techniques were employed for automatic detection and localization of pneumothorax in [60]. For the classification part, firstly the lung regions were extracted from the CXR images after which features were extracted using the Uniform Local Binary pattern (ULBP) method. Support Vector Machine (SVM) with RBF kernel was used as classifier. The second proposed method was basically the segmentation of the abnormal areas of the lung from the CXR images. The extraction of lung region was carried out after removing the background and noise in the CXR images, later the textural analysis was performed using Local Binary Pattern in order to define the smooth and complex regions. Moreover the rib boundaries were highlighted using Sobel Edge detection method. The final image was the segmented abnormal region with pneumothorax in form of a black-and-white mask. The dataset was obtained from a Chinese Hospital containing 84 CXRs. For training purpose 36 normal CXRs and 22 CXRs with pneumothorax were used while the evaluation was done on 16 normal and 10 pneumothorax CXRs. For classification purpose 82.2% accuracy was obtained while for the second proposed method 85.8% accuracy was achieved on test set.

In [61] detection and localization of pneumothorax was performed using local dataset containing 1003 CXRs obtained from University of Washington Medical Centre. Three different deep learning techniques including CNN, Multiple Instance Learning and Fully convolutional networks were experimented. For CNN, a Pretrained ResNet-50 architecture was fine trained using single channel input with resolution of 448 x 448. MIL was used for both classification and localization using the previously trained ResNet50 model as patch classifier. For FCN, a U-Net model was deployed in order to detect the location of pneumothorax. The three models achieved AUC of 96%, 93% and 92%. The authors concluded that CNN achieved best result for classification purpose while MIL and FCN perform well for identifying the location of pneumothorax.

### 4.2. Multiple pathologies related research

Many authors have contributed in the field of automatic detection of multiple thoracic pathologies including pneumothorax. The problem of multiple pathologies classification is solved either as multi-class or multi-label classification problem. Multi-class classification problems are those in which the labels are dealt independently and the classifier assign any one of the N labels to each sample, where N refers to the number of labels/classes in a dataset and N ∈ {1…..∞}. Multi-label classification problems are those in which a sample can be annotated with more than one labels [62]. This section summarizes the valued work done for the spontaneous detection of thoracic pathologies.

#### 4.2.1. Classification

A 121-layered CNN architecture was proposed in [22] for the detection of 14 pathologies from the chest radiographs which was named as CheXNet. It was basically a 121 layered Dense Convolutional Network, the weights of the model trained on ImageNet dataset were used for initializing the proposed network. CheXNet was initially proposed for pneumonia detection but later it was extended to the detection of other thoracic diseases. The trained model generated a vector with binary labels and the length of vector corresponded to the total number of classes in the classification problem, i.e. 14. The NIH Chest X-ray-14 dataset was used for this study which was divided in such a manner that 70% was used for training, 10% for validation and remaining 20% was declared as test set, while considering that same patient’s radiograph was not present in more than one split of the data. Binary cross-entropy was used as loss function with Adam optimizer and input resolution of 224x224. The model performance was evaluated in terms of AUC with achieved value of 88.87% for the detection of pneumothorax. Although heat maps were generated in order to visualize the area of pathology however the accurateness of localization was not reported.

Automatic detection of 14 thoracic diseases was carried out while solving the multi-label classification problem in [63]. Two approaches were used to tackle the multi-label classification problem, first one was training a CNN model with two loss functions including binary cross entropy (after transforming the multi-label problem as binary classification using Binary Relevance approach) and Pair Wise Error Loss. Second one was designing a six leveled cascaded architecture especially designed to tackle the multi-label classification problem that took advantage from the training strategies of boosting methods. The base CNN architecture utilized in this research was DenseNet161 which was initialized with He norm for training purpose. During training Stochastic Gradient Descent (SGD) optimizer was used with learning rate of 0.1 and momentum set to 0.9. NIH Chest X-ray-14 dataset was used for experimentation which was randomly split into 80% training and 20% testing set. In case of pneumothorax detection, best results were obtained from the cascaded model with reported AUC of 85.94% on the test set.

In [64], multi-label classification problem was studied using openly available NIH Chest X-ray14 dataset. The interdependency among labels was leveraged using LSTMs. An encoder and decoder based RNN was proposed for automated detection of thoracic diseases form the CXR image. The encoder part of the proposed classifier was based on the DenseNet. The images were resized to 256 x 256 and instead of transfer learning method, the model was trained from scratch. Maximum Log-Likelihood Estimated (MLE) was optimized during model training. The dataset was divided such that 70% was used for training, 10% for validation and 20% as test set while considering all the 14 classes. Upon testing, the model attained overall AUC of 79.8% while on pneumothorax 84.1% AUC was achieved.

In [65], a three-branch attention guided convolutional neural network (AGCNN) was proposed for the automatic diagnosis of 14 thoracic diseases including pneumothorax. The model was trained using global and local images on two baseline models, ResNet50 and DenseNet121. The global images were trained using transfer learning method and heat maps were generated which were used to crop the identified area of pathology in a CXR image. The cropped area was treated as local image and hence local region from all the training samples were obtained. These local regions were then trained on CNN architecture. Finally the last pooling layer for both the models (for training global and local images) were concatenated. Since the research dealt with multi-label classification problem, so final output was a vector with binary values, indicating the presence or absence of respective classes. NIH Chest X-ray-14 dataset was used which was randomly split in such a way that 70% was used for training, 10% for validation and 20% as test set. For training and testing purpose the images were resized to 224 x 224 and model was trained using SGD optimizer. The CNN architectures were initialized with the weights of the models trained on ImageNet dataset. The proposed AGCNN achieved best results with DenseNet121 as baseline CNN architecture, with reported AUC of 87.1% on whole test set and 92.1% on pneumothorax detection.

A multi-label classifier was designed with the aim to automatically diagnose the 14 thoracic pathologies in [66]. NIH Chest Xray-14 dataset was used for experimentation purpose. Two different protocols for data split were considered. Firstly the data was randomly divided into 70% training, 10% validation and 20% testing set, and secondly the official data split provided by the NIH was used. ResNet50 was used as base architecture and three main domains were covered in this research: (1) the effect of transfer learning with and without fine tuning and training a model from scratch. (2) Effect of high resolution of input image. (3) Effect of using non-image features like gender, age and view point along with the image features in order to train the classifier. During training, single channel X-ray images were used with resolution of 448x448. For the random split the best achieved results for pneumothorax detection were obtained when pre-trained ResNet50 was fined tuned on the selected dataset with achieved AUC of 87.0%. In case of official split, 81.9% AUC was achieved with ResNet50 architecture. However best overall results were achieved when model was exclusively trained on the selected dataset with non-image features incorporated during training. Although GradCAM was used to generate heat maps in order to visualize the area of pathology however the precision of localization methodology was not reported.

In [67], the automatic differentiation between normal and abnormal CXRs was the main motive. The main idea behind this study was to differentiate between normal and abnormal CXRs based on the synthetic image produced by the proposed model. If the trained model performed poor while reconstruction of the input CXR, then the CXR was declared as abnormal one, i.e. the CXR belong to one or more thoracic pathologies. The proposed framework was similar to Generative adversarial networks (GAN) and it comprised of three main modules. (1) Auto encoder (based on U-Net architecture) consisting of a 5 layered encoder and 5 layered decoder to generate the synthetic image. (2) A decoder (Convolutional Neural network) which was utilized for adversarial training in order to generate more realistic synthetic CXR images. (3) Another Encoder which was programmed to increase the consistency between the latent spaces of the two encoders. For model training and evaluation, the resolution of images was set to 64 x 64 and Adam optimizer was used. The experiments were performed using NIH Chest Xray14 dataset and training was done using only the normal CXRs (with no pathology). The training data consisted of 4479 CXRs while for validation purpose 849 normal and 857 CXRs with any pathology were used. In the test set, there were 667 abnormal and 677 normal CXRs present. The CXRs with at least one of the 14 thoracic diseases including pneumothorax were declared as abnormal. Four types of loss functions including image reconstruction loss, adversarial learning loss, encoding consistency loss and feature map consistency loss were utilized in this study. The model training was done keeping input size of 64x64. Upon testing, the model achieved AUC of 84.1% for discriminating between normal and abnormal CXRs.

In [68], automatic detection of 14 thoracic diseases including pneumothorax was performed using transfer learning method. The CNN architecture chosen for experiment purpose was DenseNet12, which was trained on the selected dataset using ImageNet weights as network's initializer. The last fully connected layer with 14 channels was removed and a 1024D feature vector was obtained from the trained model. As preprocessing step, the CXR images were resized to 224 x 224. Adam optimizer was used and initial value of learning rate was set to 0.001. The 1024D features obtained from the DenseNet121 architecture were fed to Logistic Regression classifier for training purpose. Since the research aim was to study multi-label classification so three different problem transformation techniques were experimented including Binary Relevance (BR), Classifier Chain (CC) and Label Power set (LP). Two openly available CXRs dataset were used in this study, i.e. 112,120 CXRs belonging to NIH chest X-ray-14 dataset and 134,327 CXRs from CheXpert dataset. Each dataset was randomly divided such that 80% was used for training and 20% was kept aside for testing purpose. The results proved that BR performed best of all the three multi-label approaches experimented in this research. For the detection of pneumothorax, the model achieved AUC of 92.9% and 86% on the test set of NIH Chest Xray14 dataset and CheXpert respectively.

A novel framework for detection of multiple thoracic diseases was proposed in which DenseNet121 was used as baseline CNN architecture in [69]. The main aspect of this framework was that it incorporated spatial information of the disease along with exploiting the high resolution of CXR image. It was found that training a model with higher resolution of input and providing location information of the pathology yield better classification results. The model training was carried out using two openly available datasets including NIH Chest Xray14 dataset and PLCO dataset [70]. In this research the labels were dealt independently instead of dealing the problem as multi label classification problem. 70% of data from each dataset was used for training the model while 10% was used for validation and remaining 20% was used as test set. Patient-wise split was considered while allocating training, validation and test set. The training was done by combining the training sets of both the datasets however the labels were dealt independently, i.e. the images with the same label in both the datasets were treated as different classes. On the other hand the testing was performed separately on the test set of NIH and PLCO datasets. In case of NIH dataset both the official train-test split and random split of dataset was considered while evaluating the proposed model. Weighted cross entropy loss was used in order to cater the class imbalance problem. The image resolution was kept 1024 x 1024 and Adam optimizer was used with initial value of learning rate set to 0.001. The model achieved overall AUC of 80.7%, 84.1% and 87.4% on NIH official split, NIH (random patient wise split), and PLCO test set respectively. The PLCO data didn't contain pneumothorax CXRs however for NIH dataset, AUC value of 84.6% and 87.2% was obtained for pneumothorax detection on test set of NIH official split and random split respectively.

In [71], deep learning model was trained in order to detect four pathologies including pneumothorax, nodule, airspace opacity and fracture. Two different datasets were deployed in this study including a local dataset (DS1) obtained from different hospitals of India and second being NIH Chest Xray14 dataset. The DS1 contained total of 759,611 CXRs while 112,120 CXRs from NIH dataset were used in this study. The x-ray images were randomly split into training, validation and testing set while considering that same patient's CXRs were present in only one set, either training, validation or test set. Certain selection criterion were followed prior to training the model. The test set of DS1 contained 1818 while 1962 CXRs were present in NIH Chest x-rays test set belonging to the four pathologies, out of which 94 and 195 CXRs belong to pneumothorax from DS1 and NIH respectively. Training was done using Pretrained Xception model by combining the training sets of DS1 and NIH, while performance was evaluated on the test sets of two datasets separately. The model performance was evaluated in terms of AUC with obtained values of 95% and 94% for DS1 and NIH respectively.

A deep learning based framework was proposed for the automatic diagnosis of multiple thoracic diseases from the CXR images in [72]. Xception model was used in this multi-label classification problem. The research was carried out using Random Sample of NIH Chest X-ray dataset which was randomly divided into 75% training and 25% testing set. The images were resized to 128x128. Adam optimizer with learning rate of 0.001 was chosen in this study. The model was trained for 28 epochs with binary cross entropy loss function. All the 14 classes present in the dataset including pneumothorax were used in this experiment, however the No-finding samples were excluded during model training. The AUC achieved for pneumothorax detection on the test set was 54% and the accuracy achieved by whole test set was reported to be equal to 88.76%.

Deep learning techniques were studied in [73] with the aim to detect the presence of 14 thoracic diseases including pneumothorax as a multi-label classification problem. The presented work comprised of two parts, firstly a five layered CNN architecture was proposed which was trained from scratch with random initialization of weights and secondly pre-trained VGG-16 architecture was fine-tuned on the selected dataset. Random Sample of NIH Chest X-ray (RS-NIH) dataset was used in this research, which was randomly split into 80% training 20% for testing purpose. Out of 80% data, 20% was declared as validation set for evaluation of model during training. The images were resized to 128x128 and Adam optimizer was used with initial value of learning rate set assigned to be 0.001. Since this study dealt with multi-label classification problem so categorical cross-entropy function loss function was used. GradCAM method was used to generate heat map in order to visualize the area of pathology however the pathology localization was not the main goal so only classification model performance was reported in terms of accuracy. Upon testing, the model achieved overall accuracy of 83.67% and 97.8% for the two different trained models.

Bharati [74] proposed a new framework (VDSNet) for the automatically identifying the presence or absence of thoracic pathology. The proposed framework was the combination of VGG16 architecture, data augmentation and spatial Transformer network STN. The CXR images were converted from RGB to gray prior to model training and the pixel values of images were normalized so that the pixel values lie within the range [0-255]. In addition to images, the Meta data like age, gender and view position of the CXR were also used for training purpose. Along with accuracy which was the main performance measure in this research, $F_{0.5}$ was also calculated. Two datasets including Random Sample of NIH Chest X-ray (RS-NIH) dataset and NIH Chest X-ray-14 dataset were used. Although both datasets have 14 classes each, however this research dealt it as a binary classification problem, i.e. the model was trained in order to detect if the CXR is normal or abnormal (any thoracic pathology). The proposed model achieved 70.8% accuracy and 64% $F_{0.5}$ score on RS-NIH while 73% accuracy and 68% $F_{0.5}$ score was achieved on NIH dataset.

In order to automatically diagnose the presence of 14 different chest diseases MIMIC CXR dataset was used consisting 473,064 CXR images in [75]. CNN was trained with different experimental setting. The main task was to find the effectiveness of using different view position CXRs. DenseNet121 was used as base architecture with little modification. 70% of the data was used for training, validation was done using 10% of data while remaining 20% was used for testing purpose. Each subset was further divided based on the view positions i.e. lateral view position, anteroposterior (AP) posteroanterior (PA) view positions. The CNN model was trained separately on these three viewpoints using ImageNet weights and input size was set to 512 x 512. A DualNet was also proposed in which both frontal view positions (AP and PA) along with Lateral view CXRs were used for training purpose. When evaluated, the results showed that DualNet performed better in terms of AUC with achieved value of 72.1% on whole test set and 62.5% for pneumothorax detection.

An automatic detection model was proposed with the aim to detect five different thoracic pathologies including pneumothorax in [76]. A deep CNN was proposed which was similar to AlexNet. The data was acquired from local hospital's Radiology Information System (RIS). One of the motive of this research was to explore the effect of DCGAN on classification performance. The original dataset contained only 4013 samples of pneumothorax, 15781 Normal, and remaining 36,626 samples belonged to four other pathologies. The resolution of input images was kept as 256 x 256 and the batch size of 128 was used during model training. The CNN model was trained for 100 epochs and Adam was chosen as optimizer with initial value of learning rate set to 0.001. The model output was a vector with five probability values corresponding to each pathology. The evaluation was done 1000 CXRs with equal contribution from each class. It was found the overall performance was enhanced when DCGAN generated CXRs were added in the training set. For pneumothorax diagnosis, the model trained with original dataset achieved accuracy of 57.99% however the accuracy value was increased to 88.84% when the model trained with original data along with DCGAN generated CXRs was used.

### 4.2.2. Classification and Localization

In [26], along with proposing a very large chest radiographs database (NIH Chest X-ray 14 dataset), experiments were performed in order to classify and locate multiple pathologies in a CXR images. The dataset contained 112,120 frontal view CXRs which was divided into 70% training, 10% validation and 20% testing set. Multi-label classification was performed using different pretrained CNN architectures including AlexNet, GoogleNet, ResNet50 and VGG16, with ResNet50 performing best of all. Various types of loss functions including Hinge Loss, Euclidian Loss and Cross Entropy Loss (CEL) were experimented to tackle the multi-label classification task. Since CEL performed best of all, it was modified a little bit to solve the class imbalance problem and the new loss function was named as weighted CEL (W-CEL). 79.93% AUC was achieved for the automatic detection of pneumothorax in the test set. Moreover localization of the pathologies was also performed by generating bounding boxes which were obtained by applying an ad-hoc thresholding based method on the heat maps obtained from the trained DCNN models. The performance of pathologies localization was evaluated for only 983 images for which 1600 annotated bounding boxes were present and the accuracy for pneumothorax localization was 17.35% with Average False Positive Rate of 52.43%.

A framework was proposed in [77] for the detection and localization of multiple thoracic pathologies by combining the concepts of multi-resolution and MIL (multi-instance learning). The LogSum-Exp pooling function was parameterized with a learnable lower bound adaptation (LSE-LBA). This resulted in more robust diagnosis along with high resolution saliency maps for localization of pathology. The proposed model consisted of ResNet architecture which was used to down sample the input image and then the image resolution was preserved using DenseNet. Training and evaluation was performed using NIH Chest X-ray14 dataset with the official data split. The proposed model achieved AUC of 76.1% on whole test set, while for pneumothorax detection and localization, the model achieved AUC of 80.5% and Dice coefficient score of 3.9% with lower-bound adaptation ($r_0$=0).

A novel approach for identification and localization of chest diseases form the radiographs was proposed by Li et al [78]. The main goal of this study was to lessen the dependency of the location information for localization task, since the acquisition of pixel-level annotations for each pathology is an expensive job. The images were first processed with ResNetV2 after which patch slicing layer was used to resize the features extracted from the convolutional neural network using max pooling or bilinear interpolation. These resized images were then trained on a fully convolutional recognition network which generated two outputs, the label prediction and an image with location information. NIH Chest Xray14 dataset was used with all the 14 labels including pneumothorax. For experimentation the images with bounding box (880 images) were separated from the images with no-bounding boxes (111,240 images) and these two sets of images were named as annotated and unannotated. The dataset was split such that 70% of data was declared as training set, 10% was used as validation and remaining 20% was kept aside for testing purpose. When evaluated, the model achieved AUC value of 87% and IoU score of 63% for pneumothorax detection and localization respectively.

## 5. Performance Metrics

There are many different performance metrics which can be used to evaluate the classification models. In case of class-imbalance datasets, AUC is preferred over other performance measures [79]. Moreover in literature mostly authors have reported the superiority of their proposed methods in terms of AUC. However few researchers have calculated other performance metrics including accuracy, sensitivity, precision and specificity. Area under Receiver Operating Characteristics (AUC) is defined in terms of true positive (tpr) and false positive rate (fpr) [80] and the formulae for these two are given below.

$$tpr = \frac{Total\ No\ of\ correctly\ classified\ positive\ samples}{Total\ No\ of\ Positive\ \ samples} \tag{1}$$

$$fpr = \frac{Total\ No\ of\ incorrectly\ classified\ negative\ samples}{Total\ No\ of\ Negative\ \ samples} \tag{2}$$

The other performance measures can be defined with the help of a confusion matrix as shown in Table 3.

**Table 3**: Confusion Matrix Plot

| | **Predicted Class** | |
|---|---|---|
| **Actual Class** | Negative (Normal) | Positive (Pathology) |
| Negative (Normal) | TN | FP |
| Positive (Pathology) | FN | TP |

The expressions for calculating accuracy, recall/sensitivity, specificity and precision [81] are given below.

$$Accuracy = \frac{TN + TP}{TN + FP + FN + TP} \tag{3}$$

$$Recall = \frac{TP}{TP + FN} \tag{4}$$

$$Precision = \frac{TP}{TP + FP} \tag{5}$$

$$Specificity = \frac{TN}{TN + FP} \tag{6}$$

In order to prove the authenticity of a localization model, the measure of overlap between the original location information and predicted location is tried to be maximized and is calculated by means of Dice coefficient and Intersection over Union as follows [82].

$$\text{Dice coefficient} = \frac{2\,|X \cap Y|}{|X| + |Y|} \qquad (7)$$

$$IoU = \frac{|X \cap Y|}{|X \cup Y|} \qquad (8)$$

## 6. Analysis and Discussion

This section compares the existing techniques for pneumothorax detection on the basis of usability in real life and reliability along with proposing some new ideas for further research.

**Table 4**: Comparison of Existing approaches for Pneumothorax detection

| Author | Technique | Dataset | Result (%) | Domain | Data split | External testing |
|---|---|---|---|---|---|---|
| **PATHOLOGY SPECIFIC RESEARCH** | | | | | | |
| Jun [39] | Ensemble of three CNNs | NIH | AUC=91.1 | C | Random | × |
| Sze [41] | Transfer learning using CheXNet | CheXpert | AUC=70.8 | C | Random | × |
| Geva [42] | Local Binary Patter with KNN | Local | Sen=81 Spe=87 | C | Random | × |
| Taylor [43] | Multiple CNNs | Local NIH | *Local (Internal set)*: AUC= 94, Sen= 55 *NIH (External set)*: AUC=75, Sen=49 | C | Random | ✓ |
| Wang [47] | Network In Network and Histogram Equalization | Local & NIH | *Local:* AUC=98.4 *NIH:* AUC=99.06 | C | Random | × |
| Park [48] | Transfer Learning using YOLO Darknet19 | Local | *Internal*: AUC=98.4,Sen =89 *External:* AUC=90.5,Sen =63 | C | Random | ✓ |
| Mostayed [50] | Modified traditional U-Net | SIIM | DSC=76.0 | L | Random | × |
| Jakhar [51] | U-Net with ResNet backbone | SIIM | DSC=83.4 | L | Random | × |
| Viniavskyi [52] | CAM generation, IRNet, U-Net model | SIIM | DSC=76.69 | L | Official | × |
| Abedalla [53] | Two stage U-Net based framework | SIIM | *Stage1:* DSC=85.02 *Stage2:* DSC=83.56 | L | Official | × |
| Tolkachev [54] | U-Net with various CNNs | SIIM | *Stage1:* DSC=85.47 | L | Official | × |
| Groza [55] | Ensemble of LinkNet(s) | SIIM | *Stage1:* DSC=88.21 *Stage2:* DSC=86.14 | L | Official | × |
| Luo [57] | Fully convolutional DenseNet | Local | DSC=92.0 | L | Random | × |
| Ouyang [58] | Multiple segmentation models | Local | IoU=66.69 | L | Random | × |
| Chan [60] | (Classification)ULBP with SVM, (Localization) LBP with Sobel Edge Detector | Local | *Classification:* Accuracy=82.2 *Localization:* Accuracy=85.8 | C&L | Random | × |
| Gooben [61] | (Classification) Pre-trained ResNet50 (Localization) MIL, U-Net | Local | *Classification:* AUC=96 *Localization:* AUC=93 | C&L | Random | × |
| **MULTIPLE PATHOLOGIES RELATED RESEARCH** | | | | | | |
| Rajpurkar [22] | CheXNet using DenseNet121 | NIH | AUC= 88.87 | C | Random | × |
| Kumar [63] | DenseNet161, Binary Relevance, Pair Wise Approach | NIH | AUC=85.94 | C | Random | × |
| Yao [64] | Encoder and Decoder based RNN | NIH | AUC=84.1 | C | Random | × |
| Guan [65] | AGCNN using ResNet50 and DenseNet121 | NIH | AUC=92.1 | C | Random | × |
| Baltruschat [66] | Transfer Learning using ResNet50 | NIH | *Random split:* AUC=87.0 *Official split:* | C | Random & Official | × |
| | | | AUC=81.9 | | | |
| Tang [67] | Generative Adversarial Network | NIH | AUC=84.1 | C | Random | × |
| Allaouzi [68] | Transfer Learning using DenseNet121 | NIH & CheXpert | *NIH:* AUC=92.9 *CheXpert:* AUC=86 | C | Random | × |
| Guendel [69] | DenseNet121 | NIH | *Random split:* AUC=87.2 *Official split:* AUC=84.6 | C | Random & Official | × |
| Mojkowska [71] | Pre-trained Xception model | Local & NIH | *Local:* AUC=95 *NIH:* AUC=94 | C | Random | × |
| Mondal [72] | Multi-label classification problem using Xception | RS-NIH | AUC= 54 | C | Random | × |
| Prakash [73] | Own architecture, Transfer learning using VGG16 | RS-NIH | Accuracy=83.76 Accuracy=97.81 | C | Random | × |
| Bharati [74] | VGG16, data augmentation, STN | RS-NIH & NIH | *RS-NIH:* Accuracy=70.8 *NIH:* Accuracy=73 | C | Random | × |
| Rubin [75] | DenseNet121 trained on different view positions of CXRs | MIMIC | AUC= 62.5 | C | Random | × |
| Salehinejad [76] | Own network similar to AlexNet, DCGAN | Local | Accuracy=88.8 | C | Random | × |
| Wang [26] | NIH Dataset Collected, DCNNs, Heat maps | NIH | *Classification:* AUC=79.93 *Localization:* Accuracy=17.35 | C&L | Official | × |
| Yao [77] | Multi resolution and Multi instance learning | NIH | *Classification:* AUC=80.5 *Localization:* DSC=3.9 ($r_0$=0) | C&L | Official | × |
| Li [78] | ResNet50, patch slicing, FCN | NIH | *Classification:* AUC=87 *Localization:* IoU=63 | C&L | Random | ✗ |

NIH: NIH ChestX-ray-14 dataset, RS-NIH: Random Sample of NIH Chest X-ray dataset, SIIM: SIIM ACR Pneumothorax Segmentation Dataset, C: Classification, L: Localization, C&L: Classification and Localization, Sen: Sensitivity, Spe: Specificity

Note that for the papers in which binary classification was considered (i.e. presence of pathology or NO pathology, without particularly specifying any pathology) [67], [73], [74], we have reported the accuracy/AUC of the binary classifiers. While for all the papers in which more than one pathology was considered in either multi-label or multi-class classification problem, we have reported the performance for the detection of pneumothorax instead of model performance on whole test set containing multiple pathologies.

### 6.1. Comparative Analysis

- It can be easily observed from the Table 4 that for pneumothorax detection (either disease specific or multiple pathologies detection system), very less work has been done for combined diagnosis and localization of affected area. The reason might be that, before the availability of SIIM ACR pneumothorax dataset (the sole purpose of which was to contribute in the field of segmentation of the area of lung affected by pneumothorax) mostly authors have used was NIH Chest Xray14 dataset which provide bounding box information for only 980 images out of which only 98 samples belong to pneumothorax class. Moreover the other dataset like CheXpert and MIMIC-CXR-JPG mainly focused on providing image-level labels instead of pixel-level annotation. It is understandable that obtaining pixel level annotation is not an easy job however localization of a pathology especially pneumothorax is important as the treatment depends entirely on the size of pneumothorax. This problem can be solved by proposing such localization techniques which lessen the dependency on the requirement of pixel level annotation for training and later predicting the affected area of lung. Efforts have been made in order to solve this problem (as in [58], [78]), however there is still great scope of improvement in this regard.
- Although NIH Chest X-ray 14 dataset provided official split of data however many researchers have used their own split instead of official data split, this creates lack of uniformity in the training and testing data of different researches. Hence for the datasets which provide official data split, shall be used as it is, instead of random split of dataset of author's own choice in order to make fair comparison of the achieved results with the benchmark results.
- Generalization is a very important matter to be considered while proposing a framework for automatic detection of pathology. Generalizability means that a model trained and validated on a dataset obtained from same experimental setting shall perform well on the data obtained from different experimental setting. In other words a model trained on dataset A shall be able to give vigorous performance for dataset B as well. However it can be observed that in most of the researches only internal test data (i.e. test data obtained from same experimental setting as the training one) was used for evaluating the proposed model, while very few have reported the performance of the proposed framework on external test data as well [43],[48]. Hence validation of the results on external dataset shall be made in order to safely use the proposed framework in real life.

- In some of the researches although huge amount of data is used for training purpose while very small test sets were used for evaluation of model ([41], [52], [67] and [76]). However for more reliability and authenticity the evaluation shall be performed on maximum possible samples.

### 6.2. Future Scope

- Since the performance of a deep learning model is dependent on the size of the training data [83], while all the CXRs dataset available till date have very few samples of pneumothorax compared to whole dataset size, so contribution shall be made by the researchers in collecting larger dataset with greater number of pneumothorax samples.
- As most of the researchers used NIH Chest Xray14 dataset while very few have used CheXpert and MIMIC-CXR-JPG dataset, so these dataset need to be explored further for pneumothorax detection purpose.
- It is a known fact that most of the medical images datasets are imbalance in nature, i.e. contain fewer samples of pathology as compared to normal samples [84]. This might generate biased results toward majority class samples. Most of the researchers have tried to solve this issue using algorithm level approaches, e.g. adding weights in the loss function in order to give more priority to less occurring samples as in [26], [41], [57]. However a comparison of existing class imbalance approaches like the one made in [20] has never been made particularly for chest radiographs dataset. As the class imbalance approaches are domain dependent [85] so conducting such a comparison and later proposing a new technique might give surprising result and can be an interesting research topic.
- It is very clear that from the last decade the trend has been shifted form Machine Learning (ML) to Deep learning techniques (DL). Mostly, the proposed frameworks used pre-trained Convolutional Neural Networks (e.g. VGG16, ResNet, and DenseNet) for classification, and for localization purpose (e.g. U-Net). So there is great scope to engage one-self in this field. However it can be observed that for classification purpose mostly researchers have used DenseNet or ResNet, so it can be an interesting experiment to try some other and new CNN architectures like EfficientNet [86] in automating the field of automatic detection of thoracic pathologies especially pneumothorax. Similarly for the purpose of localization, there are few other and relatively new segmentation models like Feature Pyramid Network (FPN) [87], LinkNet [88] and Pyramid Scene Parsing Network (PSPNet) [89] which need to be explored.

## 7. Conclusion

Chest radiography is one of the cheapest and easy available diagnostic tool used for diagnosis of various chest pathologies including pneumothorax. The successful employment of deep learning techniques in various fields of medicine has encouraged the researchers to contribute towards automating the diagnostic process from chest radiograph. Hence many different frameworks have been proposed utilizing various machine learning and especially deep learning techniques. However a summarized overview of the existing literature for pneumothorax detection is not available so far. This paper presents a systematic literature review of the research carried out in last decade using chest radiographs for the automatic detection of pneumothorax along with highlighting some research gaps. This will guide the researchers to contribute towards filling those gaps and selecting optimal technique for further research.

## Conflict of Interest

The authors have no conflict of interest to disclose.

## Funding

No funding was received for this study.